# The realisation of fast X-ray computed tomography using a limited number of projection images for dimensional metrology

Wenjuan Sun[1], Stephan Chretien[1,2,3], Ander Biguri[4], Manuchehr Soleimani[5],

Thomas Blumensath[6], Jessica Talbott[1]

1. National Physical Laboratory, Teddington, TW11 0LW, UK
2. Alan Turing Institute, London, NW1 2DB, UK
3. University of Lyon 2, Laboratoire ERIC, 5 av. Mendès France, 69676 Bron, Cedex, France
4. Department of Applied Mathematics and Theoretical Physics, Cambridge University, Wilberforce Road, Cambridge, CB3 0WA, UK
5. Engineering Tomography Lab (ETL), University of Bath, Bath, UK
6. School of Engineering, the University of Southampton, Southampton, SO17 1BJ, UK

**Abstract**

Due to the merit of establishing volumetric data, X-ray computed tomography (XCT) is increasingly used as a non-destructive evaluation technique in the quality control of advanced manufactured parts with complex non-line-of-sight features. However, the cost of measurement time and data storage hampers the adoption of the technique in production lines. Commercial fast XCT utilises X-ray detectors with fast detection capability, which can be expensive and results a large amount of data. This paper discussed a different approach, where fast XCT was realised via the acquisition of a small number of projection images instead of full projection images. An established total variation (TV) algorithm was used to handle the reconstruction.

The paper investigates the feasibility of using the TV algorithm in handling a significantly reduced number of projection images for reconstruction. This allows a reduction of measurement time from fifty-two minutes to one minute for a typical industrial XCT system. It also enables a reduction of data size proportionally.

A test strategy including both quantitative and qualitative test metrics was considered to evaluate the effectiveness of the reconstruction algorithm. The qualitative evaluation includes both the signal to noise ratio and the contrast to noise ratio. The quantitative evaluation was established using reference samples with different internal and external geometries. Simulation data were used in the assessment considering various influence factors, such as X-ray source property and instrument noise.

The results demonstrated the possibility of using advanced reconstruction algorithms in handling XCT measurements with a significantly limited number of projection images for dimensional measurements. This work lays down the foundation for conducting fast XCT measurements without the need for instrument alteration or enhancement. Although the reconstruction time required is still considerable, various possibilities to improve this have been discussed.

# 1. Introduction to the recent development of X-ray computed tomography for industrial applications

X-ray computed tomography (XCT) has experienced rapid growth in recent years due to the significant demands of non-destructive evaluation (NDE) of advanced manufacturing, led by additive manufacturing [1-3]. The applications initially focused on imaging internal defects, such as pores and cracks. However, more and more quantitative evaluations, such as measuring the size of internal channels, the diameter of spheres, the thickness of walls *etc.,* are required. Many measurement tasks are associated with industrial and medical applications, where there is a stringent requirement about data quality and measurement efficiency.

The quality of XCT measurements can be influenced by many factors. There is an increasing number of papers concerning the factors such as thermal drift, geometric errors, and imaging artefacts. However, there has been very limited study regarding the influence of reconstruction algorithms on dimensional metrology. Most investigations were focused on the qualitative evaluation of images rather than quantitative evaluation. A recent paper discussed the evaluation of reconstruction algorithms using beads, but the investigation focused on statistical analyses over a large number of beads [4].

Filtered back-projection (FBP) based reconstruction algorithms are analytical algorithms that are still widely used due to their efficiency in reconstruction. The FBP based algorithm used for cone-beam XCTs was developed by Feldkamp, Davis, and Kress (FDK) [5, 6]. However, FBP based algorithms require a large number of projection images to fulfil the Shannon sampling theorem [7] to eliminate image artefacts [8]. FDK algorithm for cone-beam XCT systems also inevitably has cone beam effects that may cause image distortion and errors in evaluation, especially for features far from the centre of the rays along the rotation axis. The number of projection images ($n$) required is related to the detector width ($S_{detector}$) and the detector pixel size ($S_{pixel}$), see Eq 1. A typical industrial cone-beam XCT system with a detector size of 400 mm × 400 mm and a detector pixel size of 200 μm × 200 μm requires 3142 projection images. Therefore, one complete scan requires at least 52 minutes if the exposure time is 1 second per projection. This is a significant barrier when applying XCT on production lines, where the speed of measurements is of great importance to limit the cost of inspection.

$$n = \frac{\pi s_{detector}}{2 s_{pixel}} \quad \text{Eq 1}$$

Apart from the scan speed, data storage is also an issue. For a 16-bit detector, one full scan results is roughly 23.4 Gb of raw data. The total size of data is likely to be doubled when taking into consideration the reconstructed data. Therefore, FBP based algorithms are costly when a large number of data are required for recording purposes.

There are two potential solutions to reduce the scan time and data size. One potential solution is to increase scan speed physically, where a fast X-ray detector is required. However, fast detection is likely to come with a cost of a reduction in the number of pixels, low resolution and reduced image quality. Also, to achieve a reasonable image quality within a short time, an advanced X-ray source with higher flux may be required. In addition, the volume of the data collected becomes problematic.

The other potential solution is to reduce the number of projection images acquired from each scan. In the literature, this is often called "sparse-view" XCT [9]. The research around "sparse-view" XCT is predominantly for medical applications to reduce patients' exposure to radiation. But the research on its industrial applications is minimal. By reducing the number of projection images, a reduction in measurement time and data size is achieved simultaneously. This is an appealing solution to achieve fast scans without the need to replace or upgrade the existing machine's components. However, the data reconstruction based on a limited number of projections needs to be reconsidered as the FBP based algorithms cannot handle sparse-view XCT data without introducing significant artefacts. Based on the review in [10], iterative algorithms work much better with sparse-view data. To enable scans and reconstruction of limited as well as noisy projections and to reduce systematic artefacts, there are increasing efforts to develop iterative algorithms [11, 12]. Of particular interest here are total variation based algorithms [12, 13].

Total variation (TV) is a norm that measures the total adjacent pixel difference (in its discrete form) of an image. Initially introduced for image denoising [14], it can be used as a regulariser in iterative reconstruction, where it enforces piecewise smooth images, acting therefore as an artefacts remover for XCT, where most artefacts are either streak-like (beam hardening) or pixel noise (scatter). One of the first and most used TV algorithms to solve the XCT problem is Adaptative Steepest Descent with Projections Onto the Convex Set (ASD-POCS)[14], proposed for medical XCT.

In the past, the investigation of the quality of reconstruction achieved with these advanced methods was focused on image quality. Common performance measures include the signal to noise ratio, the contrast to noise ratio etc. However, the investigation of the influence of reconstruction algorithms on dimensional measurements is absent.

In the recent European Metrology Programme for Innovation and Research (EMPIR) project Advanced Computed Tomography for dimensional and surface measurements in the industry (AdvanCT), scans using circular trajectories with few or limited angles have been investigated with the aim of significantly reducing the scan time to a few minutes or less. Recent advances in the reconstruction algorithms, especially the TV-based algorithm, have been studied and enhanced during the project [12, 15].

This paper investigates the use of XCT with significantly fewer projection images for dimensional measurements in industrial applications. A significant reduction from 3142 to 60 projection images is considered, which allows the measurement time to be reduced to 1 minute from 52 minutes. The OS-ASD-POCS, a TV-based algorithm, was studied to investigate its feasibility for handling fewer projection images. A comprehensive evaluation strategy for XCT reconstruction algorithms is elaborated. Both qualitative and quantitative evaluations are considered. Reference samples used in this work are samples incorporating geometries commonly used for industrial applications. Both rotation symmetrical and non-rotational symmetrical features are included to evaluate the sensitivity of the reconstruction algorithms to systematic errors, including beam hardening and noise. In order to evaluate these impacts individually, simulation data is used to independently control different influence factors.

## 2. Reference samples

Two reference samples designed by the National Physical Laboratory (NPL), United Kingdom, were used in this study. One reference sample is NPL's mini AM measurement standard made in aluminium, see Figure 1. The reference sample has two geometries, cylinder and cuboid. They are good representations of common geometries used in industrial applications, such as bores, channels, and walls. The features also allow an investigation of the reconstruction of geometries with different orders of rotational symmetry. The reference sample also incorporates both external and internal features so that material impact can be taken into consideration. The diameter of the cylinder and the distance between surfaces of the cuboid can be used to quantify the error of the reconstruction algorithms on form evaluation.

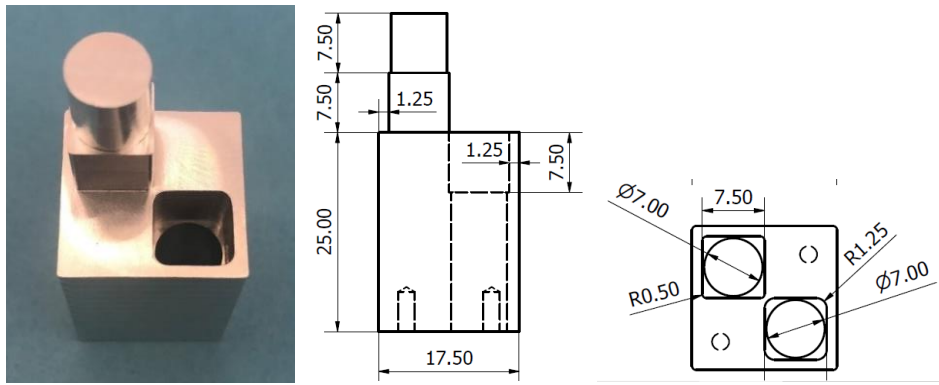

*Figure 1 NPL's mini AM reference sample in aluminium and its dimension. Left. Real AM reference sample. Middle. Specification of the sample, side view. Right. Specification of the reference sample, top view.*

The other reference sample used was the two-spheres (TS) reference sample. The reference sample has two spheres sitting on top of an aluminium post. The concept of design, *i.e.*, the centre-to-centre sphere distance, is often used to evaluate the scale factor of XCT measurements. Figure 2 shows the real object and a simplified version for simulation.

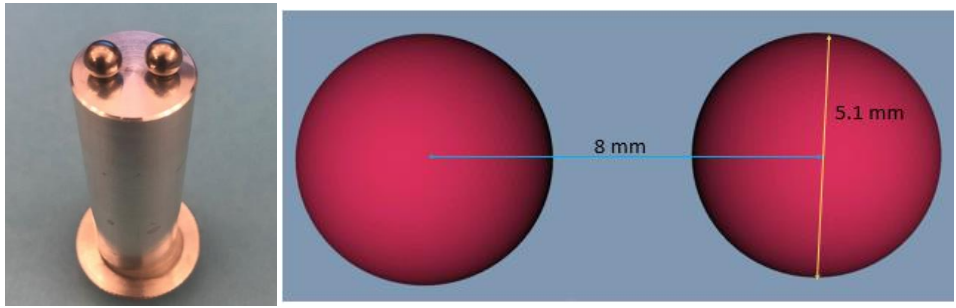

*Figure 2 NPL's TS reference sample. Left. Real TS reference sample. Right. A simplified version was used in the simulation. The spheres material considered is aluminium in the simulation.*

## 3. Methodology

### 3.1. Reconstruction algorithms

The reconstruction algorithms evaluated were available from the open-source software TIGRE (version 2.2) [16]. The software package has several reconstruction methods, but one analytical algorithm (FDK) and two iterative algorithms, ordered-subset simultaneous algebraic reconstruction (OS-SART) and ordered-subset adaptive steepest descent projection onto convex set (OS-ASD-POCS) algorithms, were selected for this work.

OS-SART is used for non-regularised reconstruction. It is a modification of the traditional simultaneous algebraic reconstruction technique (SART) algorithm [17] that uses several projections of a fixed size to update the image iteratively, instead of updating the image projection by projection. While this essentially slows down convergence per iteration, it is much faster computationally speaking and thus is preferred in real-life applications.

The OS-ASD-POCS algorithm is an ordered subset version of the well-known ASD-POCS algorithm [14]. ASD-POCS allows for the minimisation of separate convex optimisation problems, in this case, data minimisation and total variation minimisation, in an adaptative manner that ensures global convergence. ASD-POCS was proposed with SART as a data minimiser. But OS-SART can also be used as a data minimiser for ASD-POCS for faster speed, namely OS-ASD-POCS.

### 3.2. Data preparation

To thoroughly evaluate the quality of reconstruction, the study used simulation data so that various influence factors, such as beam hardening and instrument noise, can be considered separately. The simulation was performed using the aRTist 2.10 simulation software [18, 19]. To emulate common industrial XCT systems, the simulation considered a cone-beam X-ray source with a flat panel detector. The flat panel detector has a pixel array of 2000 × 2000 with a 200 µm × 200 µm detector's pixel size. A detector multisampling of 3 × 3 has been considered to eliminate errors due to discretised data in the aRTist.

The study has two parts. The first part is the study of the AM reference sample. As the AM reference sample has a cylinder and cuboid shape, which is effectively a stack of images of circles and squares, the study was focused on two-dimensional (2D) cross-section evaluation. A comprehensive study is detailed below and described in Figure 3.

- The first part of the simulation considered a monochromatic X-ray source without noise as ideal data.
- The influence factors, beam hardening and noise, were then considered. It should be noted that beam hardening is a common issue due to polychromatic X-ray sources. Therefore, the following text uses a polychromatic X-ray source to indicate the occurrence of the beam hardening effect.
- The FDK algorithm was then used to reconstruct the full (3142) projection images on both monochromatic and polychromatic data, and the results were used as benchmark data.
- The study then investigated OS-ASD-POCS algorithms for handling limited (60) projection images. OS-SART has also been evaluated as a comparison to illustrate the advantage of OS-ASD-POCS.

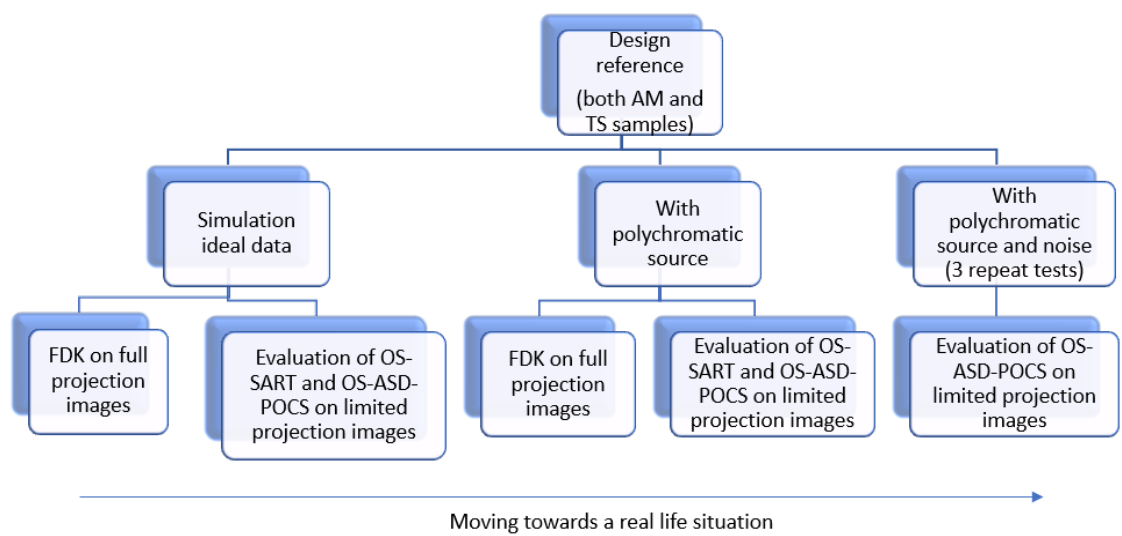

*Figure 3 Methodology of the study on the AM reference sample.*

The second part is the study of the TS reference sample. As a sphere has a three-dimensional geometry that the cross-section images vary along all three axes of the coordinate system, the study of the TS reference sample focused on three-dimensional (3D) assessment. Centre to centre sphere distance has also been studied as the other quantity to evaluate the reconstruction algorithms.

When evaluating the OS-ASD-POCS algorithm, only 60 projection images were used. For common industrial XCT systems, with a detector exposure time setting of 1 second per projection image, a scan of 60 projection images reduces the measurement time to one minute, assuming the time to rotate sample is minimum. This is to evaluate the performance of the reconstruction algorithms with an extremely limited number of projection images, which is less than 2% of a full scan of 3142 projection images for the given detector. However, in day to day operation, the number of projection images is the choice of users, depending on the requirement of the measurements.

When conducting quantitative evaluation, it is necessary to define the surface (or edge) prior to analyses. There are many edge determination algorithms available. However, this work uses the Canny edge determination algorithm due to its modest speed and accuracy [20, 21]. The algorithm was established by John Canny [22] and is widely used for surface determination.

### 3.3. Qualitative and quantitative test metrics

In this study, both qualitative and quantitative evaluations were considered. The qualitative evaluation considered the common parameters in imaging evaluation, such as the signal to noise ratio(SNR) and the contrast to noise ratio(CNR), whereas the quantitative evaluation focused on the dimension of geometries as test metrics. Apart from them, the speed of reconstruction was also included in the evaluation.

- Qualitative test metrics

SNR is an important measure of imaging quality widely used in the field of medical imaging. SNR is defined as the ratio of the mean signal from a region of interest (ROI) to its standard deviation [23], see Eq 2.

$$SNR = \frac{S}{\sigma} \qquad \text{Eq 2}$$

Where $S$ is the mean signal from an ROI. $\sigma$ is the standard deviation of the ROI.

CNR is defined as a ratio of the difference in the mean signal intensities between two image areas to the total noise of the two signal intensities [24], see Eq 3.

$$CNR = \frac{|S_a - S_b|}{\sqrt{\sigma_a^2 + \sigma_b^2}} \qquad \text{Eq 3}$$

Where $a$ and $b$ denote two areas under evaluation.

- Quantitative test metrics

Quantitative evaluation of the reconstruction algorithms was rarely considered in the literature. The quantitative evaluation in this paper is conducted by evaluating features with known dimensions, such as cylinders, cuboids and spheres. They are dimensional features commonly used in industrial applications.

When evaluating the reconstruction algorithms using the AM measurement standard, as the dimensions of features do not vary along the vertical axis, the investigation was conducted in 2D,

*i.e.*, the cross-section images of the reconstructed data. The cross-section features to be evaluated in the AM reference sample are shown in Figure 4.

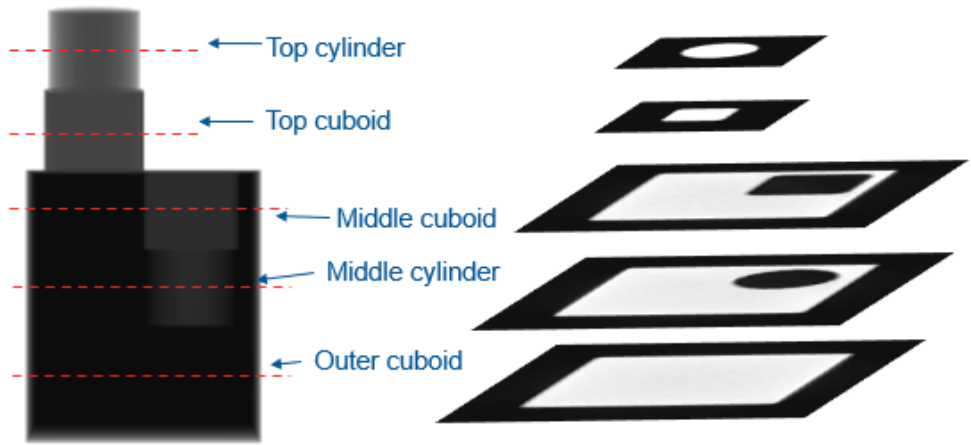

*Figure 4 Indication of the measurements on the AM measurement standards. Left. Side view of the sample tomographically. Right. Cross-section images. From top to bottom, the cross-section features evaluated are outer circle (OCir), outer square (OSqu), inner square (ISqu), inner circle (ICir), large outer square (LOSqu).*

The cross-section image was then processed with the edge determined for quantitative evaluation. The edge data of the related feature was then evaluated using the least-squares fitting of lines or circles [25, 26]. Examples of features determined are illustrated in Figure 5. The combination of the outer and inner geometries in different shapes and sizes allows a comprehensive evaluation. The reference value is given in Table 1. For square features, different colour markers shown in Figure 5 indicate the edge determined. The notation used in Table 1, for example, $OSqu\_dis_x$ is the mean edge distance between the 'left' and the 'right' edge of the outer square shown in Figure 5 and $OSqu\_dis_y$ is the mean edge distance between the 'top' and the 'bottom' edge.

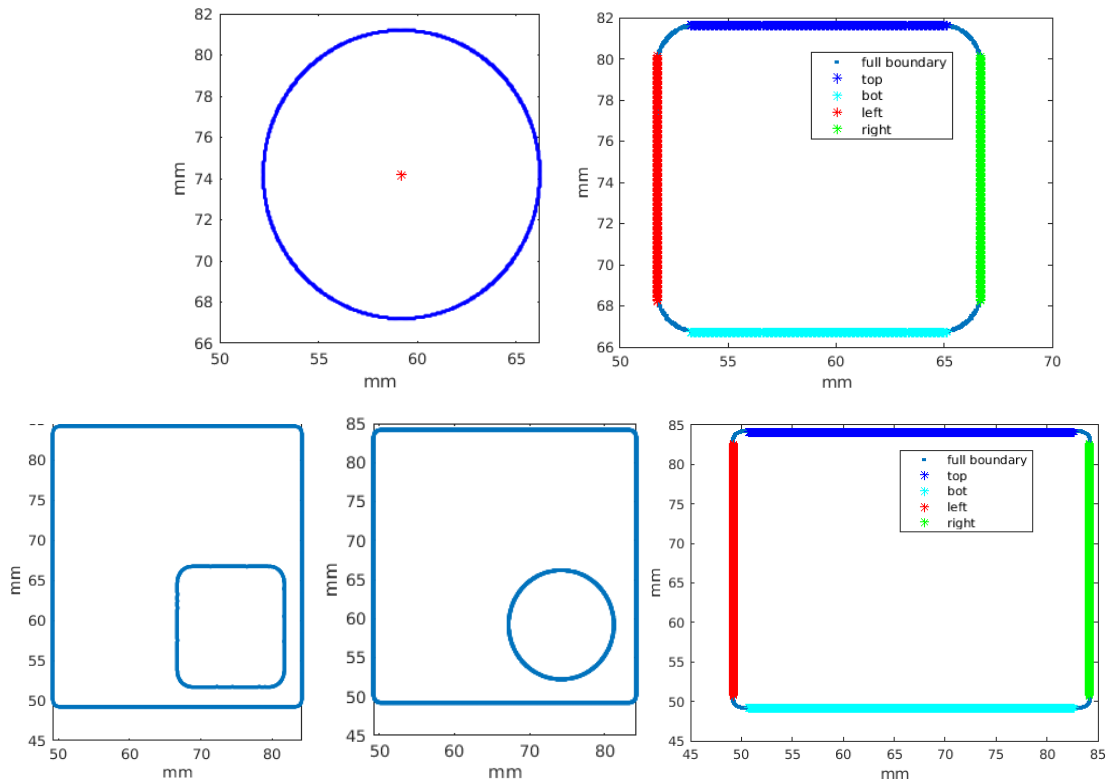

*Figure 5 Illustration of feature determined. Top left. OCir determined. Top right. OSqu determined. Bottom left. ISqu and outer determined. Bottom middle. ICirl determined. Bottom right. LOSqu determined. Colour markers indicate the edge data determined of square features.*

*Table 1 Reference value of the AM measurement standard used in the simulation.*

| Features | Parameters | Nominal：mm | Explanation of notation |
|---|---|---|---|
| Cross-section of the outer cylinder | OCir_R | 7 | The radius of the outer circle |
| Cross section of the top cuboid | OSqu_dis$_x$ | 15 | Mean distance between the edge 'left' and 'right' of the outer square |
| | Osqu_dis$_y$ | 15 | Mean distance between the edge 'top' and 'bot' of the outer square |
| Cross-section of the inner cuboid | ISqu_dis$_x$ | 15 | Mean distance between the edge 'left' and 'right' of the inner square |
| | ISqu_dis$_y$ | 15 | Mean distance between the edge 'top' and 'bot' of the inner square |
| Cross-section of the inner cylinder | ICir_R | 7 | The radius of the inner circle |
| Cross-section of the bottom cuboid | LOSqu_dis$_x$ | 35 | Mean distance between the edge 'top' and 'bot' of the large outer square |
| | LOSqu_dis$_y$ | 35 | Mean distance between the edge 'top' and 'bot' of the large outer square |

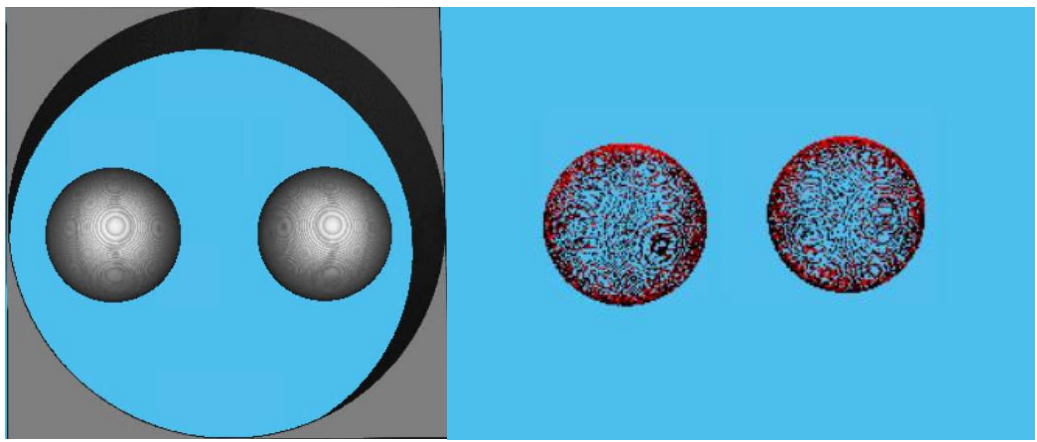

*Figure 6 Reconstruction of the TS reference sample. Left. Volumetric data. Right. Point clouds of the spheres.*

When evaluating the reconstruction algorithm using the TS reference sample, the focus was on the 3D evaluation. Both dimensions of the spheres and the centre to centre sphere distance were evaluated. The quantitative evaluation was based on the nonlinear least-squares sphere fitting algorithm [27]. An example of sphere determination is shown in Figure 6. The reference value is given in Table 2.

*Table 2 Reference value of the TS reference sample*

| Parameters: mm | Sphere diameter | Centre to centre sphere distance |
|---|---|---|
| Values | 5.1 | 8 |

Apart from the qualitative evaluation and quantitative evaluation, speed is also an important factor considered. To have a balanced view on the performance of reconstruction algorithms, consideration was given to the time to conduct measurements, load, save, and reconstruct data.

## 3.4. Computing unit

The computing unit plays an important role in the reconstruction. The computing facility utilised for this study is high-performance computing on the NPL's High Performance Computer (HPC). The reconstruction software and evaluation were conducted in Matlab 2021a. The specification is shown in Table 3.

*Table 3 Specification of NPL's HPC facility*

| **NPL's HPC facility** |
|---|
| 16 compute nodes, each node comprising: |
| • 2 x 16 CPUs, Intel Skylake 2.6 GHz |
| • 192 GB RAM (6 GB per CPU) |
| 1 GPU node with 2 x NVIDIA P100 GPU |

## 4. Data analyses and results

This section presents the evaluation results using the AM reference sample. The simulation considered the source to object distance of 400 mm, and the source to the detector distance of 1200 mm. The magnification was 3 and the effective pixel size (detector pixel size/magnification) was 66.7 μm × 66.7 μm. The investigation focuses on cross-section images of the reconstructed data.

### 4.1. Results of the AM reference sample

As introduced in section 3, the results of FDK on full projection images on both monochromatic and polychromatic X-ray source data were used as benchmarks. The OS-SART and OS-ASD-POCS algorithms on 60 projection images were then compared to the benchmarking results.

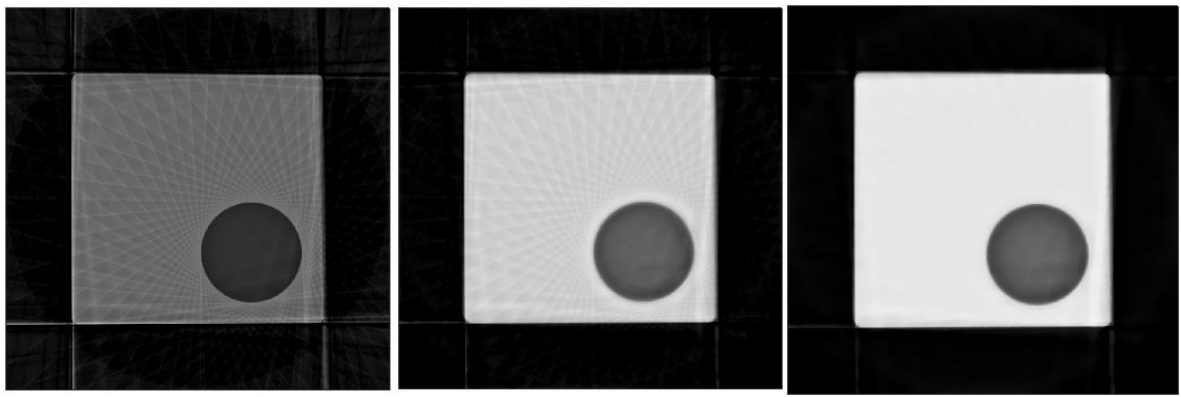

*Figure 7 Reconstruction based on 60 projection images using different reconstruction algorithms. Left. FDK algorithm, Middle. OS-SART algorithm. Right. OS-ASD-POCS algorithm.*

To illustrate the challenges when handling a limited number of projections and also to demonstrate the effectiveness of TV-based algorithms, FDK has also been applied to the test using 60 projection images. From Figure 7, it can be seen that the result of the FDK algorithm has significant artefacts. The OS-SART algorithm result has shown a reduction of artefact, but the problem is still noticeable. In contrast, the image reconstructed using the OS-ASD-POCS algorithm has demonstrated a good image quality in terms of surface uniformity and feature edge preservation.

*Table 4 Test results of the AM reference sample using simulation data, with monochromatic or polychromatic X-ray source, without noise.*

| Influence factor | | Nominal value | Difference between measured results and nominal value | | | | | |
|---|---|---|---|---|---|---|---|---|
| | | | Monochromatic source | | | Polychromatic source | | |
| | | | Full projection | 60 projection images | | Full projection | 60 projection images | |
| Feature parameters | | | FDK | OS SART | OS ASD POCS | FDK | OS SART | OS ASD POCS |
| Outer cylinder | OCir_R:mm | 7 | 0.000 | 0.001 | 0.000 | 0.002 | 0.004 | 0.003 |
| | OCir_SNR | 12.767 | 0.000 | -2.722 | -2.235 | -0.517 | -1.199 | -0.632 |
| | OCir_CNR | 12.558 | 0.000 | -2.894 | -2.345 | -0.390 | -1.342 | -0.769 |
| Outer cuboid | OSqu_dis$_x$:mm | 15 | -0.067 | 0.001 | -0.004 | 0.011 | 0.008 | 0.003 |
| | OSqu_dis$_y$:mm | 15 | -0.067 | -0.024 | -0.023 | 0.006 | 0.002 | -0.018 |
| Inner cuboid | Isqu_dis$_x$:mm | 15 | 0.053 | -0.004 | -0.003 | 0.055 | -0.003 | -0.004 |
| | Isqu_dis$_y$:mm | 15 | 0.053 | 0.016 | 0.033 | 0.055 | 0.018 | 0.032 |
| Inner cylinder | ICir_R:mm | 7 | 0.001 | 0.000 | 0.001 | 0.000 | 0.000 | 0.000 |
| Outer cuboid (large) | OLSqu_dis$_x$:mm | 35 | 0.000 | 0.025 | 0.006 | 0.000 | 0.034 | 0.030 |
| | OLSqu_dis$_y$:mm | 35 | 0.000 | 0.059 | 0.021 | 0.000 | 0.110 | 0.137 |

The comparison of using OS-SART and OS-ASD-POCS algorithms on limited (60) projection images and FDK on full projection (3142) images are shown in Table 4. Both simulations using a monochromatic source and a polychromatic source were considered. The evaluation was based on the cross-sections of the cylinder and cuboid features (*i.e.* circle and square, respectively).

When using FDK with full projection images, the results of the outer cylinder and inner cylinder, for both monochromatic and polychromatic data, have small errors compared to the nominal value of the reference sample. However, the results of the square features have an error of between zero and a voxel size, which is due to the pixelisation error during the edge determination process.

When evaluating OS-SART and OS-ASD-POCS with 60 projection images, from a quantitative perspective, the test on the inner and outer circle features has a good agreement with the FDK algorithm and nominal values. Both algorithms run 60 iterations. When evaluating the square features, the errors of OS-SART and OS-ASD-POCS are generally greater than the errors of the circular features. The errors of the square features were due to combined pixelisation error and image quality. In most cases, these errors are less than one voxel, apart from the large outer square.

Especially with a polychromatic X-ray source, the impact of beam hardening has an inevitable impact on the edge determination of square features. The internal feature causes a further impact on the beam hardening effect and results in large errors on the outer square features (*i.e.*, the OLSqu_dis$_y$ has a larger error than the results of OLSqu_dis$_x$).

For the qualitative evaluation (including SNR and CNR), as there are no reference values, results of FDK on the full projection data using a monochromatic source were used as nominal values. In all cases, SNR and CNR of OS-ASD-POCS are closer to the results of using FDK with full projection data compared to OS-SART, indicating that OS-ASD-POCS are superior to OS-SART in terms of imaging quality.

Evaluation of square features using XCT is deemed to be less accurate than the analyses of circular features. It can be noticed from Figure 7 that when reconstructed with a significantly limited number of projection images, the resulting image using FDK has severe imaging artefacts. The artefact consequently may affect the edge determination and hence the quantitative evaluation. This is likely due to the thresholding selection of the Canny edge determination, which is not robust in handling images with local gray value variations due to artefacts [28]. However, this is an area that requires more investigation. Typical errors caused are such as detecting fake edges, missing edges, or deforming edges. They generally occur around square features in this study. Figure 8 illustrates the influence of low image quality and its impact on edge determination of square features.

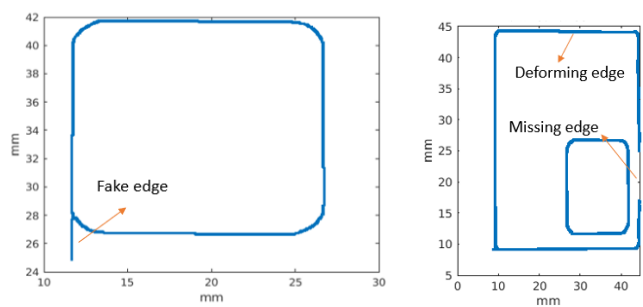

*Figure 8 Typical errors in resulting edges. Left. Wrong edges. Right. Deforming and missing edges.*

In the quantitative evaluation, fake edges can be eliminated if they are not part of the features to be evaluated. Missing edges also have little impact as the distance between edges of square features is based on averaged edge position. However, the deforming edges cannot be neglected, where the errors in the results inform the quality of quantitative evaluation. These errors may occur at the corner of the square features and the part with complicated internal structures (see Figure 8 Right).

Based on the results in Table 4, it can be seen that the OS-SART and OS-ASD-POCS algorithms are able to handle extremely small numbers of projection images. However, OS-ASD-POCS algorithm is more capable of handling significantly limited projection images and demonstrated a better result, especially in qualitative evaluation compared to OS-SART.

Further to the investigation of data with monochromatic and polychromatic sources, the influence of noise has been considered. The noise factor was simulated using the default noise setting in the aRTist software. The influence was considered in addition to the polychromatic source. Three repeat simulations have been conducted. The results of OS_ASD_POCS using 60 projection images are shown in Table 5.

*Table 5 Evaluation of the OS_ASD_POCS algorithm using polychromatic source and noise simulation data.*

| Feature parameters | | Full projection, mono with no noise | A polychromatic source with noise OS_ASD_POCS using 60 Projection images | | | | |
|---|---|---|---|---|---|---|---|
| | | FDK | 1 | 2 | 3 | mean | std |
| Outer cylinder | OCir_R:mm | 7.0002 | 0.003 | 0.003 | 0.003 | 0.003 | 0.000 |
| | $SNR_{cir}$ | 12.7667 | -0.659 | -0.606 | -0.655 | -0.640 | 0.030 |
| | CNR | 12.5576 | -0.794 | -0.743 | -0.788 | -0.775 | 0.028 |
| Outer cuboid | $OSqu\_dis_x$:mm | 14.9333 | 0.008 | 0.000 | 0.005 | 0.004 | 0.004 |
| | $OSqu\_dis_y$:mm | 14.9333 | -0.023 | -0.008 | -0.013 | -0.015 | 0.008 |
| Inner cuboid | $Isqu\_dis_x$:mm | 15.0528 | -0.001 | 0.000 | -0.006 | -0.002 | 0.003 |
| | $Isqu\_dis_y$:mm | 15.0528 | 0.019 | 0.023 | 0.034 | 0.025 | 0.007 |
| Inner cylinder | ICir_R:mm | 7.0012 | 0.001 | 0.000 | -0.001 | 0.000 | 0.001 |
| Outer cuboid (large) | $OLSqu\_dis_x$:mm | 35 | 0.031 | 0.030 | 0.029 | 0.030 | 0.001 |
| | $OLSqu\_dis_y$:mm | 35 | 0.133 | 0.056 | 0.084 | 0.091 | 0.039 |

Table 5 presents the reconstruction using OS_ASD_POCS from 60 projection images with the consideration of noise. The error of the averaged value from the circular feature is in the order of micrometre level, and the standard deviation is also at a micrometre level. Similar to the analysis of the monochromatic and polychromatic data, the errors and standard deviation of the analysis of square features are greater than the errors of the circle features.

Apart from qualitative and quantitative evaluation, the other important factor considered here is time. To have a balanced view of the overall efficiency of the reconstruction algorithm, the consideration has been given to the measurement time, image loading time, reconstruction time and finally, the time to save the data.

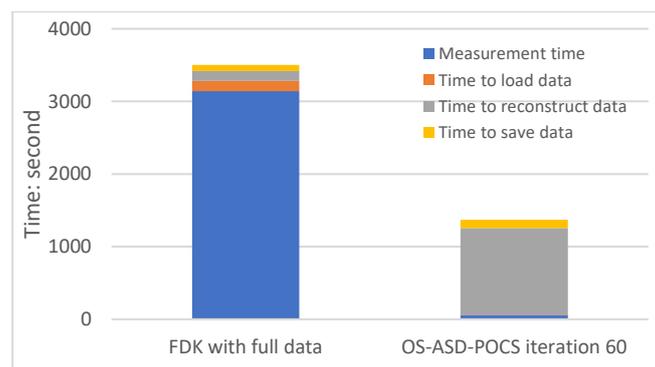

*Figure 9 Comparison of time to obtain and process XCT data using FDK on 3142 projection images and OS-ASD-POCS algorithms on 60 projection images. The detector range considered is 1400 × 800 pixels to improve the efficiency of the test. The input of OS-ASD-POCS is a 1400 × 800 × 60 data, with 60 iterations and reconstructed into an 800 × 800 × 1400 data.*

Figure 9 is the comparison between FDK using full projection images and OS-ASD-POCS using 60 projection images. With the consideration of measurement time, loading time, reconstruction, and final data saving, the OS-ASD-POCS has a great advantage. However, the reconstruction time is significant compared to FDK. Also, the case illustrated in Figure 9 was based on the part of the detector of 1400 × 800 pixels, 60 projections, and the reconstructed data size was 800 × 800 × 1400. When using OS-ASD-POCS to reconstruct a 2000 × 2000 × 2000 volume requires approximately 230 seconds per iteration, the total time cost can be extremely high with a large number of iteration. However, the time of reconstruction can be potentially improved with:

- The optimisation of the number of iterations.
- The optimisation of the size of output data.

It is not known from the literature how these can be optimised. The following sections will investigate this further.

## 4.2. Influence of iterations

The OS-ASD-POCS algorithm is an iterative algorithm and the time spent is heavily related to the number of iterations.

Figure 10 Left illustrates the evaluation of the OS-ASD-POCS algorithm using the root-mean-squared errors (RMSE). Reconstructed data using FDK with full projection images have been used as the reference in the calculation of RMSE. It can be noticed that the RMSE value decreased to a smaller number after two iterations. The reconstruction was started with an initial image value set to zeros.

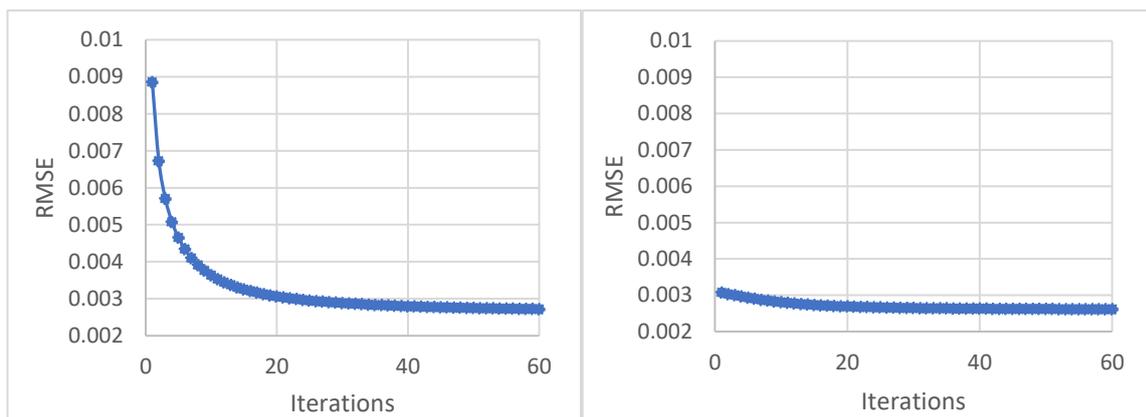

*Figure 10 RMSE of the reconstruction using OS-ASD-POCS based on simulation data using monochromatic source. Left. Image value initialised to zero. Right. Initial value estimated using FDK.*

*The reconstruction can be improved by having a better estimated initial value using algorithms, such as FDK or conjugate gradient least squares algorithm (CGLS).*

Figure 10 Right shows the improvement by using the FDK as initial estimates, which demonstrates a potential saving on the number of iterations using the FDK result as the initial estimate. When using FDK as an initial estimate, the RMSE value at 1 iteration of 0.003068 is compatible with OS-ASD-POCS results of at 26 iterations (0.003063) without an initial estimate.

The impact of iterations on both quantitative and qualitative evaluation is in shown Figure 11. The simulation data was the data set with a monochromatic X-ray source. As in the case of the previous comparison, nominal values and FDK results were used as references. The data showed continuous improvement of results in terms of qualitative evaluation. The results of circle features showed gradual improvement, but the improvement is not significant after 30 iterations. The changes in results in the evaluation of square features are more significant. However, the results are sensitive to edge determination, as discussed above.

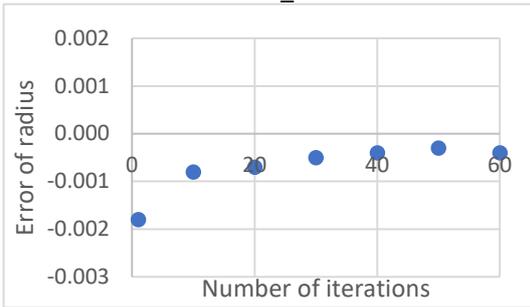
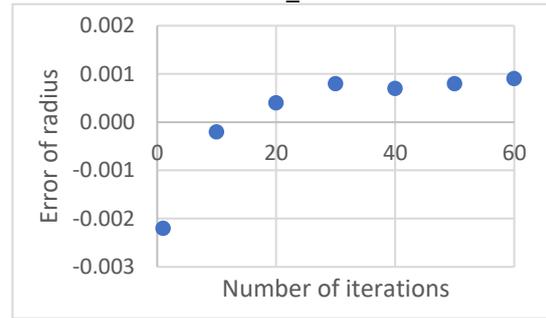
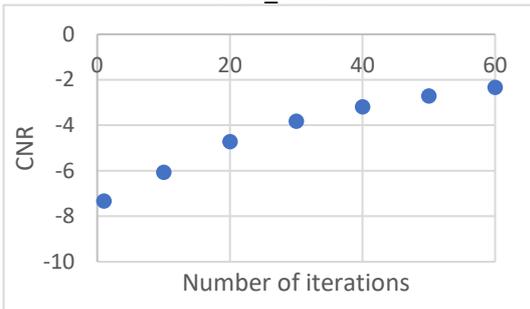
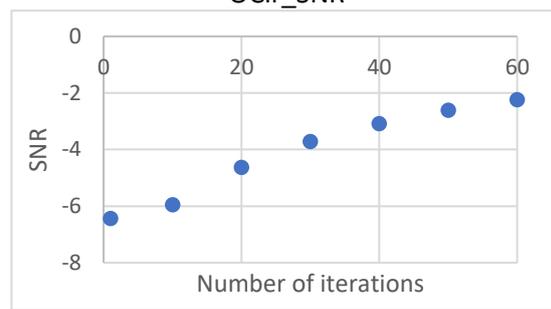
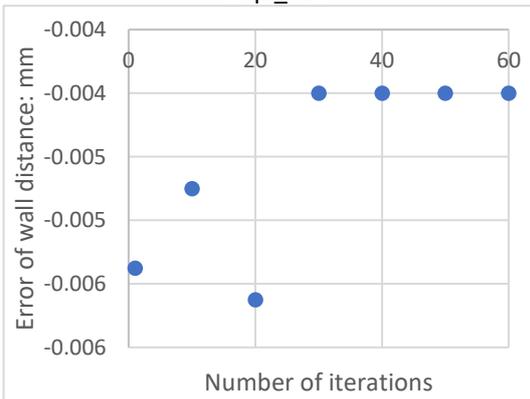
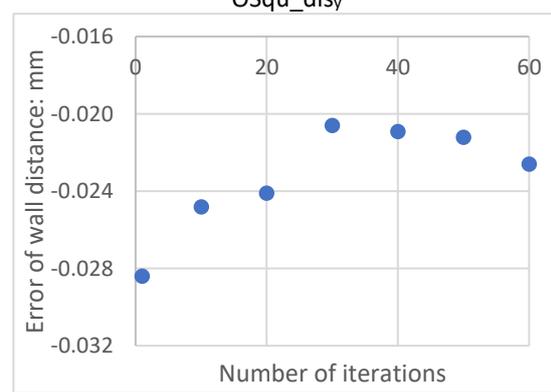
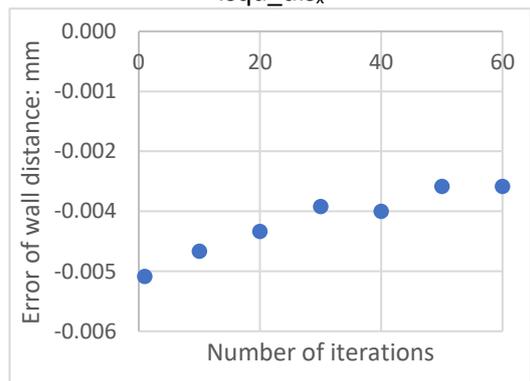
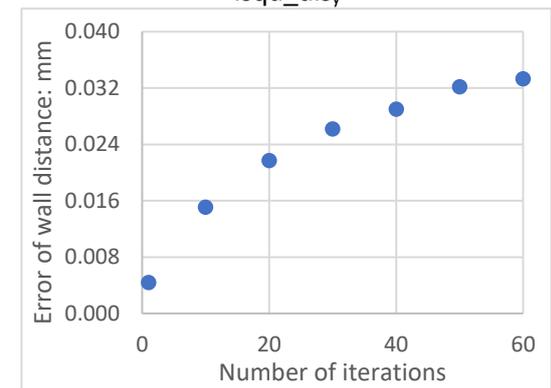

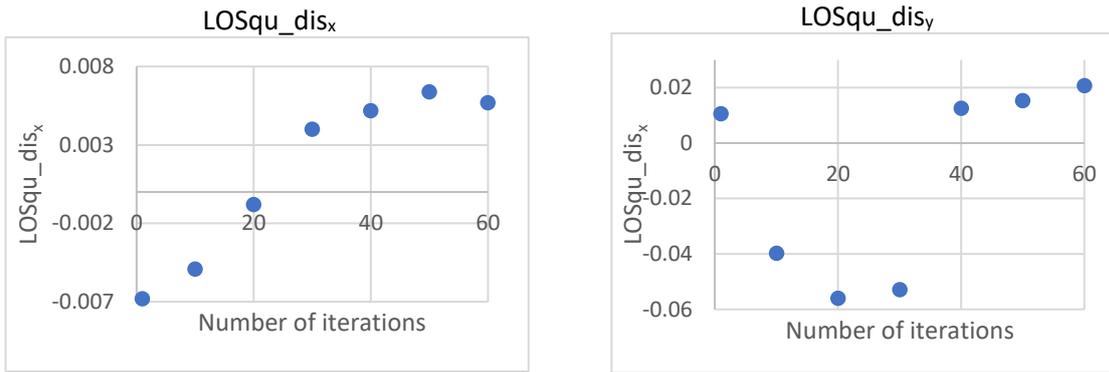

*Figure 11 Plot of errors of different features with different iterations against reference values. The simulation data was the data set with a monochromatic X-ray source. Same as the previous comparison, nominal values and FDK results were used as references.*

### 4.3. Evaluation of voxel size of reconstructed data.

The other factor to consider in optimising the process is the voxel size in the reconstructed data. In the case of using a full projection data set, the common process is to reconstruct a data set using the effective detector pixel size. However, with a limited number of projections, there may not be a need to sustain such a small voxel size. For example, at the magnification of 3, the effective detector pixel size is 66.7 µm for a detector pixel size of (200 µm)$^2$. However, with only 60 projection images, Eq 1 suggests a pixel size of 10.5 mm and an effective pixel size of 3.5 mm.

In reality, it is not ideal to use a 3.5 mm voxel size as it is very close to the size of features observed. However, it is worth investigating whether there is a reduction potential.

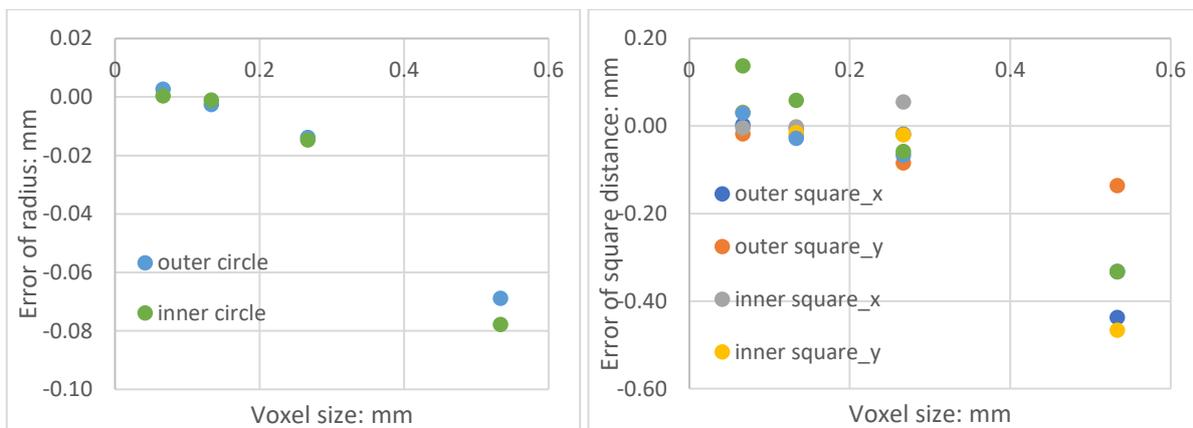

*Figure 12 Impact of sampling spacing using a monochromatic source with 60 projection images. Left. Quantitative evaluation of the radius. Right. Quantitative evaluation of square features. Distance between walls.*

Figure 12 presents the results of the evaluation using different voxel sizes in the reconstructed data. The impact of increasing the sampling spacing is very small when the voxel size increases two times, *i.e.*, (133 µm)$^3$. Even with four times of increase (266 µm)$^3$, the increase of error is much smaller than the effective detector pixel size of 66.7 µm. The error becomes significant when there is eight times increase. Figure 13 shows the time for reconstruction with different voxel sizes. It also shows a great time reduction when voxel doubles and quadruples, but no significant reduction with further increase.

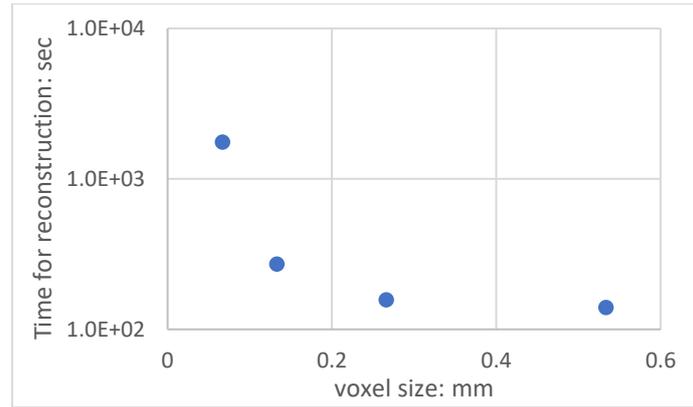

*Figure 13 Time of reconstruction using OS-ASD-POCS algorithm with 60 projection images with a monochromatic source. Projection data on the detector has been trimmed to 1400 × 800 to improve the efficiency of the test.*

Based on the results above, it is possible to improve the efficiency of the OS-ASD-POCS algorithm by one order of magnitude. When conducting a real measurement task, the parameters of the algorithm should be optimised for a specified measurement task.

## 5. Evaluation using the two-spheres measurement standard

When evaluating the reconstruction algorithms in three-dimensional space, the TS reference sample was used. The nominal value of each sphere's diameter is 5.1 mm, and the centre to centre distance between the spheres is 8 mm. The simulation considered a monochromatic X-ray source. The magnification was set at 10. With the detector pixel size 200 μm × 200 μm, the effective pixel size was 20 μm × 20 μm. The reference reconstruction was conducted using the FDK algorithm in the TIGRE software package. The surface was determined using the Canny edge determination algorithm, and the surface geometry was estimated using the nonlinear least-squares fitting algorithm [27]. The reconstructed volumetric data and the surface determined are illustrated in Figure 6, and the results are shown in Table 6.

*Table 6 Results of the simulation of the TS reference sample. Evaluation of OS-ASD-POCS algorithm was conducted using 60 iterations.*

|  | No of images | Reconstructed voxel size: μm³ | Time to load: second | Time to reconstruct : second | Error of Diameter of sphere 1: mm | Error of diameter of sphere 2: mm | Error of sphere distance : mm |
| --- | --- | --- | --- | --- | --- | --- | --- |
| FDK | 3142 | 20 | 324.5 | 30.3 | -0.0008 | -0.0008 | 0.0001 |
| OS-ASD-POCS | 60 | 20 | 1.6 | 473.1 | -0.0015 | -0.0011 | 0.0000 |
| OS-ASD-POCS | 60 | 40 | 1.7 | 146.2 | -0.0043 | -0.0033 | 0.0004 |
| OS-ASD-POCS | 60 | 80 | 2.0 | 105.8 | -0.0133 | -0.0119 | 0.0000 |
| OS-ASD-POCS | 60 | 160 | 1.7 | 90.2 | -0.0465 | -0.0474 | -0.0032 |

It should be noted that the detector range has been truncated to 800 × 400 pixels to save time on analysis. The reconstructed data for FDK is then 800 × 800 × 400 voxels. The reconstructed data size for OS-ASD-POCS is reduced correspondingly according to the voxel size.

The result of the TS reference sample is similar to the results of the cross-section analysis of the AM reference sample. The OS-ASD-POCS algorithm can handle measurement data with limited projection images, and the results are close to the nominal value. With a significantly limited number of projection images, the results of OS-ASD-POCS are still in close agreement with the results of FDK using full projection images. Even with 4 times increase in reconstructed voxel size, *i.e.* $(80\ \mu m)^3$, the error of sphere diameter is still less than the original voxel size of $(20\ \mu m)^3$. The error of the sphere centre distance is much small compared to the sphere diameters. Even with 8 times of increase of reconstructed voxel size, the error is 3.2 µm.

# 6. Discussion and conclusion

An extensive study has been conducted with both quantitative and qualitative evaluation of an advanced TV-based reconstruction algorithm, OS-ASD-POCS, in handling XCT measurements with a significantly reduced number of projection images. The quantitative evaluation considered various geometries in both 2D, and 3D and the qualitative evaluation considered SNR and CNR parameters. The evaluation was based on simulation data so that influence factors could be controlled.

In the comparison between FDK, OS-SART and OS-ASD-POCS, OS-ASD-POCS algorithm demonstrated the advantage of TV algorithms in handling an extremely limited number of projections, where surface edges are better preserved compared to OS-SART and FDK. When quantitatively evaluating the OS-ASD-POCS algorithm with a limited number of projections, it is noted that the OS-ASD-POCS algorithms had a good agreement with reference value when evaluating sphere centre distance. The errors on sizes (sphere and circle diameters) are slightly more than centre to centre distance but still close to reference values. When dealing with surfaces with non-rotational symmetric features (*e.g.* lines, planes), more errors are due to the combined influence of reconstruction and surface edge determination.

The evaluation in this study was conducted using only 60 projection images, which is less than 2% of the full number of projection images. Therefore, a typical industrial XCT system can reduce the scan time from 52 minutes to 1 minute. In the production line, this means a significant reduction in the cost of measurements. It also reduced the cost of data saving. Although the time to reconstruct using the OS-ASD-POCS is long, in the paper, a few possibilities have been discussed to overcome the issue. Including optimisation of the iteration process and the voxel size of reconstructed data.

It is important to note that the reconstruction times reported in this work are implementation and hardware dependent. TIGRE, albeit fast for reconstruction, has a modular design and is multi-purpose. An implementation of the algorithm for particular hardware and problem size would incur a considerable time improvement. Similarly, TIGRE can accelerate the reconstruction almost linearly [29] to up to 8 GPUs in a single CPU, given the right hardware. This work is executed in a 2 GPU machine, meaning a 4-fold reduction in computational cost can be achieved from a purposedly built computer compared to the times reported in this work. The availability of such a computer would be reasonable to expect in an industrial production pipeline. In practice, it is important to know the measurement requirement in terms of speed, errors and uncertainties so that the measurement and data process can be optimised to suit the measurement tasks.

**Acknowledgement**

This work was funded by the EMPIR programme co-financed by the Participating States and from the European Union's Horizon 2020 research and innovation programme. This work was also funded by the UK Government's Department for Business, Energy and Industrial Strategy (BEIS) through the UK's National Measurement System programmes.